# The Sun's displacement from the galactic plane from spectroscopic parallaxes of 2400 OB Stars

B. Cameron Reed

Department of Physics

Alma College

Alma, MI 48801

e-mail: reed@alma.edu

ph: (989) 463-7266

fax: (989) 463-7076

**ABSTRACT**

The Sun's vertical displacement from the galactic plane, $Z_{\odot}$, is determined model-independently from 3457 spectroscopic parallax distance-estimates for 2397 OB stars within 1200 pc of the Sun. The result, $19.5 \pm 2.2$ pc, agrees well with various other recent determinations. The distribution of average stellar z-values as a function of galactic longitude shows a slight sinusoidal dependence with an amplitude of about 26 pc.

# 1 Introduction

The Sun's vertical displacement from the galactic plane, $Z_\odot$, is an important parameter in galactic-structure models and has been the subject of a number of investigations over the years; a selected sample of results is summarized in Table 1. The most recent is that due to Joshi (2005), who examined the distribution of interstellar extinction towards several hundred open clusters within 5º of the galactic plane. In an analysis of the displacement from the plane of maximum absorption as a function of galactic longitude he found that the absorption has a sinusoidal dependence with an offset of 22.8 ± 3.3 pc, a figure he interpreted as $Z_\odot$ and which is consistent with the results of many other investigators. It should be noted that Table 1 is not exhaustive; Humphreys & Larsen (1995) summarize a number of other determinations of $Z_\odot$ dating from 1974, with, in most cases, results similar to those listed in Table 1.

Joshi's result for $Z_\odot$ derives from a fit to only 7 points (see his Fig. 9), and the majority of the results listed in Table 1 are model-dependent in various ways. Consequently, it was thought that a fresh determination of $Z_\odot$ which invokes as few assumptions as possible might be worthwhile.

In view of their concentration to the galactic plane and their small vertical scale height the obvious stellar candidates from which to estimate $Z_\odot$ are solar-neighborhood OB stars. The *disadvantage* of OB stars is that distances estimates to individual stars can be highly uncertain (~50% uncertainty is not uncommon) and so a large sample is required to get a meaningful average. Within the last few years two papers directed at determining $Z_\odot$ based on OB stars have appeared. In an earlier paper (Reed 1997; R97) I estimated $Z_\odot$ based on assigning average absolute magnitudes to a large sample (up to ~ 12,000) of OB stars with known photographic magnitudes that lay within 10º of the galactic plane and approximately 4 kpc of the Sun and assuming a barometric-type extinction model. The result ($Z_\odot$ ~ 10-12 pc) must be regarded as crude, however, as the stars were assigned absolute magnitudes on the basis of coarse objective-prism classifications that designated them as one of the three types OB$^+$, OB, or OB$^-$. Later, Maíz-Apellániz (2001) used *Hipparcos* trigonometric parallaxes for 3383 O-B5 stars with $|b| > 5°$, determining $Z_\odot$ = 24.2 or 25.2 pc depending on whether one adopts a self-gravitating isothermal or Gaussian disk plus parabolic-halo stellar distribution model. While both of these papers utilize large samples, both also make various model assumptions.

For some years I have been developing databases of published UBVβ photometry and MK spectral classifications for OB stars. These databases were originally restricted to stars cataloged in the Case-Hamburg galactic-plane surveys but were subsequently expanded to be an all-sky effort that now incorporates over 18,000 stars (Reed 2003). Instructions on accessing my catalog, data files, and supporting documentation are available at `othello.alma.edu/~reed/OBfiles.doc`. "OB stars" here is taken to mean main-sequence stars down to temperature class B2 and more luminous ones down to temperature class B9. At this writing (July 2005) my databases incorporate over 38,000 photometric observations and nearly 23,000 spectral classifications. This very extensive assembly of data for the OB stars opens up the opportunity of reexamining $Z_\odot$ on the basis of a large, model-independent sample. This is the purpose of the present work.

The selection of the sample is described in Sect. 2, with particular attention to explaining how sources of spectral classifications were judged for quality. Results are given in Sect. 3.

## 2     Sample Selection

Since the present work depends on deriving distances via spectroscopic parallaxes, isolating the best-quality classifications is essential. The approximately 23,000 classifications in my database are drawn from nearly 500 sources. As might be expected, these classifications derive from photographic and electronic material obtained with a variety of instruments employing various dispersions and wavelength coverage. All sources were examined and assigned a classification quality code from A to E with A being the highest. In brief, code A designates classifications deriving from observational material that adhered to the original MK criteria of wavelength coverage ~ 3500-5000Å, dispersion 60-125 Å/mm, and resolution 1-2Å. Code B designates classifications that nominally adhere to these criteria but which are suspected of possibly being of slightly lower quality, such as objective-prism spectra where obtaining the optimum exposure for each star can be difficult. Code C designates classifications derived from material that was obtained for classification purposes but which failed to meet the MK criteria in some way, for example, high-dispersion objective-prism or thin-prism classifications. Code D designates classifications deriving from material failing to meet the MK criteria and which was obtained for some other purpose such as radial velocity or abundance studies, and code E is a catch-all category reserved for cases where authors give little or no detail on their instrumental system and so whose classifications could not be meaningfully assessed. Only code A, B, or C classifications are used in this work. While my original intent had been to apply some weighting scheme to these codes this proved unnecessary as over 80% of the adopted spectroscopic parallaxes derived from A-quality classifications (see Sect. 3).

A program was written to merge my photometric and classifications files and to output a list of highest-quality useable classifications for all stars, with "useable" meaning a classification with both full temperature and luminosity components. In some cases lower-quality classifications were selected where nominally higher-quality ones lacked luminosity classes. If more than one highest-quality classification was found for a given star each was retained and treated separately. A second program was written to read in the output of the first and assign absolute visual magnitudes and intrinsic B–V colors for the selected classifications from the calibrations of Turner (1980) and Schmidt-Kaler (1982), respectively (interpolating linearly across either or both of temperature and luminosity class where necessary) and to compute distances via spectroscopic parallax in the usual way. The interstellar absorption is assumed to be given by

$$A_V = [3.3 + 0.28(B-V)_o + 0.04 E_{B-V}] E_{B-V} \qquad (1)$$

where $(B-V)_o$ is the intrinsic color and $E_{B-V}$ the color excess.

Since we are interested in determining the average of the stellar z-values, strict completeness of the sample to some limiting distance is not terribly crucial provided we do not otherwise unduly discriminate against some range of z-values. An examination of the results of the programs described above revealed that the number of stars for which distances could be computed begins to decline for magnitudes fainter than V ~ 10. Now, the intrinsically faintest members of the sample are B2 V stars, for which $M_V = -2.2$. If we conservatively assume galactic-plane extinction of 1.5 mag/kpc (see R97), a B2 V star of apparent magnitude V = 10 would be at a distance of about 1200 pc. To avoid discriminating against low-z valued stars, i.e., those near the galactic plane, I adopt this distance limit for the purpose of determining $Z_\odot$. This choice has two further advantages: (i) it is large enough to render fairly inconsequential any effects due to Gould's belt while (ii) being small enough to avoid incorporating any confounding effect due to the galactic warp, which makes a distance of closest approach to the Sun of ~ 6 kpc (Reed 1996).

## 3   Results

A total of 3457 classifications for 2397 separate stars yielded distance estimates within 1200 pc, with the numbers of (A, B, C)-quality classifications being (2888, 397, 172). In computing $<z> = -Z_\odot$, distance estimates for stars with more than one estimate were weighted as the inverse of the number of estimates for that star in order that all stars are weighted equally. The data are available at `othello.alma.edu/~reed/N=3457.dat`.

The weighted mean for these 3457 distance estimates gives

$$Z_\odot = 19.5 \pm 2.2 \text{ pc},$$

where the error limit is the standard error of the weighted mean, computed as the weighted standard deviation divided by the square root of the sum of the weight s.

This present result falls comfortably at about the middle of the limits of those given in Table 1. Curiously, it is a little greater than those deriving from analyses of galactic infrared dust emission [Cohen (1995); Binney et al. (1997)], which may hint that the OB stars are somewhat more vertically displaced than the material from which they formed. Conversely, it lies to the low end of Joshi's (2005) estimate based on open clusters. This too is not surprising upon considering that some of the clusters in his sample are as old as 10 Gyr whereas the greatest main-sequence lifetime for stars in the present sample is ~ 20 Myr (B2 dwarfs).

Figure 1 shows average z-values for stars in 10-degree-wide bins of galactic longitude *l* as a function of longitude. The error bars are the standard errors as described above but computed separately for each longitude bin. A very scattered sinusoidal trend is evident. Performing a least-squares fit via the Simplex method (Caceci & Cacheris 1984) to a curve of the form

$$Z = A \sin(l + \phi) - Z_\odot \qquad (2)$$

yielded $(A, \phi, Z_\odot) \sim (26 \text{ pc}, 50°, -16 \text{ pc})$. In his analysis of the height above or below the plane of the maximum differential absorption, Joshi (2005) found very roughly similar results: $(A, \phi, Z_\odot) = (41 \text{ pc}, 36° -23 \text{ pc})$. The maximum upward displacement of the OB-star "plane" relative to the b = 0 plane occurs at $l \sim 45°$. The Simplex method used here does not return meaningful error estimates for the fit parameters but an idea same can be had by examining how they fluctuate upon varying initial estimates input to the program. Against reasonable changes in the initial estimates they are in fact quite robust, varying by no more than about $\pm$ 1-2 pc in A and $Z_\odot$ and about $\pm 2°$ in $\phi$.

If Figure 1 really does reflect a tilting of the OB-star plane relative to the galactic plane, the effect is slight: an amplitude of 26 pc at a distance of 1200 pc implies a tilt angle of about 1°.2. It is worth emphasizing that this effect has nothing to do with the Gould belt, a much more local phenomenon restricted to the brighter B stars (Lesh 1968).

Given the various assumptions underlying the results summarized in Table 1 and the inevitable errors attendant the spectroscopic parallaxes utilized in the present work, more detailed interpretations are not warranted. However, it is gratifying to see that the present *model-independent* results are good accord with the various *model-dependent* results summarized in the Table and with the general trend of interstellar extinction as derived from open clusters.

Figure Caption

Figure 1

Average z-value as a function of galactic longitude in bins of longitude of width $10°$. The error bars are standard errors of the mean for each bin. The curve is the least-squares fit sine curve described by equation (2).

Table 1.

Selected recent determinations of $Z_\odot$

| Reference | $Z_\odot$ | Sample |
|---|---|---|
| Conti & Vacca (1990) | 15 ± 3 | WR stars (N = 101) within 4.5 kpc of Sun |
| Cohen (1995) | 15.5 ± 0.7 | IRAS point-source counts + point-source sky model |
| Humphreys & Larsen (1995) | 20.5 ± 3.5 | Galactic-pole star counts (N ~ 10,000) plus Bahcall-Soneira galaxy model |
| Méndez & van Altena (1998) | 27 ± 3 | Solar-neighborhood reddening model plus star counts |
| Binney et al. (1997) | 14 ± 4 | COBE/DIRBE surface-brightness analysis; double-exponential disk + power-law bulge |
| Reed (1997) | ~10-12 | OB stars with $|b| < 10°$; averaged $M_B$ values for rough OB classes; assumed extinction model |
| Chen et al. (1999) | 27.5 ± 6.0 | COBE/IRAS-based extinction model |
| Maíz-Apellániz (2001) | 24.2 ± 1.7 | *Hipparcos* parallaxes for 3382 O-B5 stars ($|b| > 5°$) within ~ 350 pc, plus distribution model |
| Joshi (2005) | 22.8 ± 3.3 | extinction analysis for ~ 600 open clusters with $|b| < 5°$ |

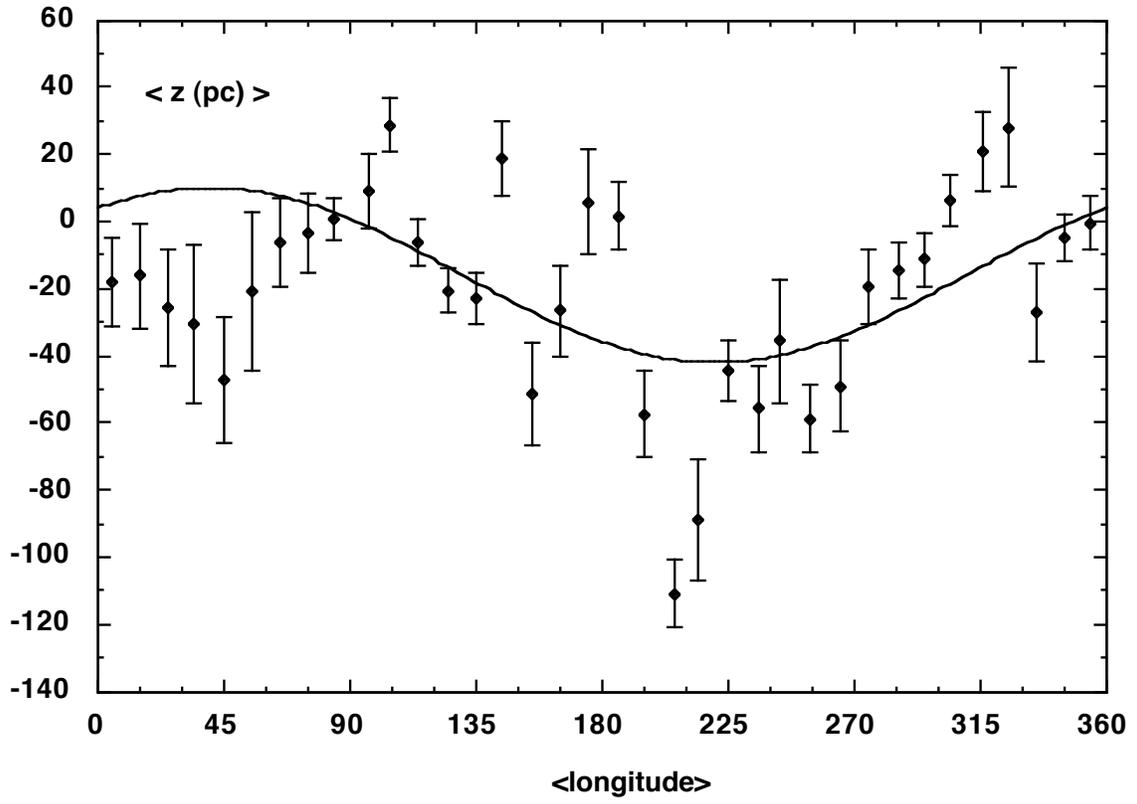

Figure 1